# Security Threats in Prepaid Mobile

[1]Arshiya Begum, [2]Mohammed Tanveer Ali

[1] Computer Science Department, King Khalid University,
Abha, KSA
*arshiyabegum@ymail.com*

[2] Network Specialist, Computer Science Corporation(CSC), Hi-Tech City,
Hyderabad, Andhra Pradesh, India
*mohdtanveerali@gmail.com*

**Abstract**
**Recent communications environment significantly expand the mobile environment. Prepaid mobile services for 3G networks enables telecommunication to sign up new users by utilizing the latest in converged billing technologies. The worldwide mobile communication market is exploding, and 50 percent of subscribers are expected to use prepaid billing. Prepaid services are driving mobile communication into emerging markets such as South America, Eastern Europe, Asia, Africa and Gulf Countries.
Prepaid phone service requires a user to make payment before calling. It is quite common to get prepaid SIM cards on every major Network. This paper discuss about various prepaid techniques, challenges and countermeasures in prepaid mobile communication system.**

## I. INTRODUCTION

Cellular telephones are increasingly become a crucial part of our daily lives. As of Year 2012, the total number of cellular phone users worldwide was 6.7 billion and this was growing at the rate of 52.49% every 12 months. In the India, the industry is signing up new subscribers at the rate of one every two seconds, putting it on track to reach 851.70 million customers sometime later this year. According to the cellular Telecommunication Industry Association(CTIA) [1], the cellular industry in the United States grew 25.3 percent in 2010, adding 93 million additional wireless subscribers, for a total of 302.1 million customers. Average usage grew 40.5 percent in 2010 to 180 minutes a month compared to 130 minutes a year ago.

In countries where mobile adoption is already high, prepaid options allow credit challenged and lower-income consumers to participate. It is very popular with first time users such as teenagers who are also early adopters of new services including Short Message Service(SMS) and games. The number of prepaid customers has rapidly grown in recent years. Prepaid customers have grown to be the largest customer group in Western Europe. Various techniques for implementing prepaid charging will be explained in the Chapter II. Additional security issues of mobile internet also discussed in chapter IV.

Mobile networks as a communications facility is viewed as a national infrastructure, because if it can be backed up with appropriate security technologies by hackers can be a victim of cyber terrorism, economic and social loss for mobile operators will be greater. We analyze the security threats in mobile networks and provide direction to solve it. In this paper, Chapter II overview of Prepaid mobile, Chapter III defines the security threats and countermeasures in mobile phones. Chapter IV introduces conclusion.

## II. BACKGROUND

*A. Prepaid in Europe and other part of the world*

Prepaid customers have traditionally been viewed as a secondary customer group, postpaid customers being the primary. They have been seen as low average revenue per user (ARPU). In addition prepaid customer group includes one very important customer segment, which is teenagers and other young people. Although this group is not very credit worthy it is fast in adopting new services.

Because operators have viewed prepaid customers as less desirable than postpaid, prepaid markets have remained untapped until the saturation point of postpaid customers has been seen in the near future.

In the Western Europe the rapid growth of prepaid customers started approximately around year 1998 [3]. Today over 60% of Western European mobile subscribers are using prepaid charging.

Prepaid and post tariffs differ significantly. Postpaid tariff plans are usually very complex with many changing factors such as monthly charge, charge per minute depending on time of day, rental of mobile device etc. Prepaid tariffs are usually a bit simpler. Often prepaid tariff consist only the charge per minute. Of course these tariffs are extremely operator dependent but on a general level I would say that prepaid tariff plan are simpler than postpaid.

*B. Features of Prepaid Mobile phones*

Following list explains few of them
- Operator gets the money before the call is made and it can invest it or at least earn interest on it.
- Operator saves in billing expenses.
- There is no credit risk.
- Some customers prefer to operate only on cash basis. Without prepaid this segment would be unreachable for operators.
- In some cultures cash transactions are preferred over credit transactions
- Some customers will never use their whole balance
- Customers who want to enjoy anonymity can also use mobile phones

*C. Prepaid in GSM networks*

The emergence of prepaid in GSM networks can be seen as normal evolutionary market development. First operators wanted to satisfy the needs of the most profitable market segment and after the growth in that segment begins to decrease the operators want to move to other segments.

There are four alternative solutions for implementing prepaid service in GSM network. These are:

- Intelligent Network(IN)
- Service node
- Hot billing
- Handset based approach

Illustration of technologies lies heavily on [4].

*1) Prepaid in IN :*

Figure 3 illustrates how intelligent network based prepaid solution is implemented in GSM networks. This and all the other illustration about technical implementation are presented . But the true complexity of these technologies lies beyond the scope of this presentation.

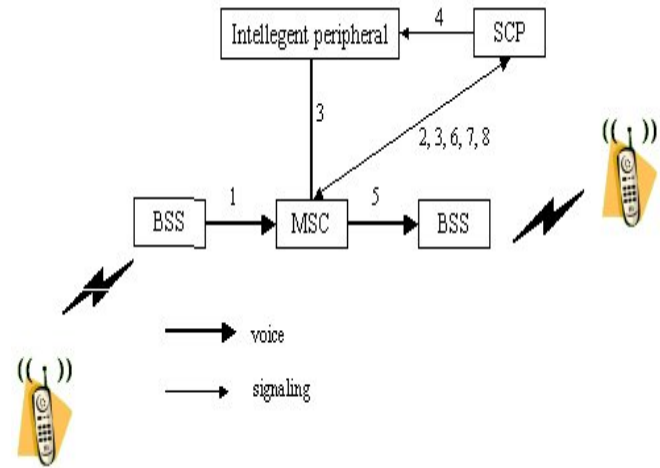

**Figure 3: Prepaid in IN**

1. Customer initiates a call
2. Mobile switching center (MSC) gets IN call setup trigger. It suspends the call and sends a message to service control point (SCP) that handles the prepaid account.
3. SCP instructs MSC to set up a voice link to intelligent peripheral. This link is used for notifications about the status of prepaid account.
4. SCP gives instructions to intelligent peripheral about account notifications.
5. SCP starts countdown timer and instructs MSC to connect the call.
6. Call terminates because countdown timer has expired or the call is completed or..
7. MSC gets IN call release trigger, sends disconnect message to SCP
8. SCP computes the cost of the call, charges the prepaid account and sends current balance and cost of the call to MSC.

*2) Prepaid by Using service note technique*

Figure 4 illustrates the implementation of service node prepaid technique in GSM networks.

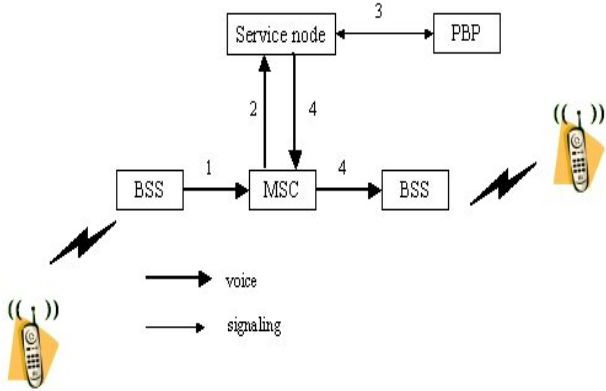

**Figure 4: Service node prepaid**

1. Customer initiates a call
2. MSC detects that the caller uses prepaid account and sets up a voice channel to service node.
3. Service node asks from the prepaid billing platform (PBP) if the call should be allowed.
4. If call is allowed, a second voice channel is established from service node through MSC to the called party

This method costs one extra voice channel compared to IN prepaid. On the other hand it is easy to implement of prepaid in GSM networks.

*3) Prepaid Using Hot Billing:*

Figure 5 illustrates the hot billing implementation of prepaid in GSM networks.

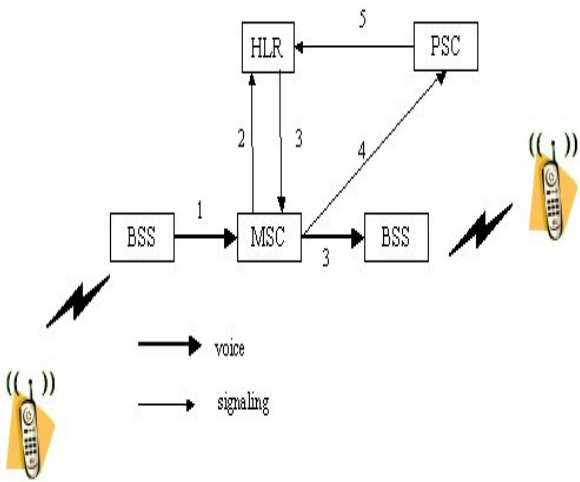

**Figure 5: Prepaid using hot billing**

1. Customer initiates a call and sends its international mobile subscriber identity(IMSI).
2. MSC asks from HLR if the service request is valid. Using IMSI HLR checks from authentication center the validity of the call.
3. HLR sends customer data to MSC and MSC connects the call.
4. When the call terminates call detail records sent to prepaid service center (PSC)
5. PSC charges the account. If the account is empty PSC notifies it to HLR and service is suspended.

In this technique the billing is not real time and the operator is exposed to credit risk of one call.

*4) Hand Set based Prepaid:*

Figure 6 illustrates the handset-based implementation of prepaid.

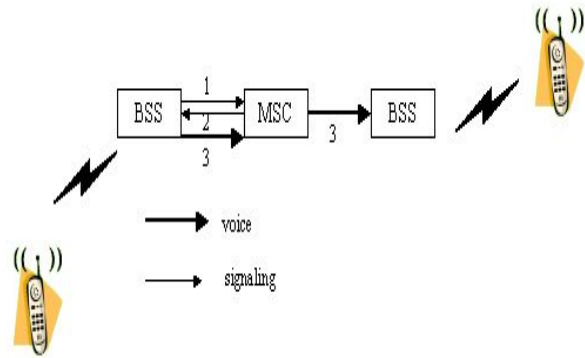

1. Customer initiates the call
2. MSC sends the pricing parameter to mobile station. Mobile station uses these parameters for decrementing the account
3. MS acknowledge the parameters and the call is connected
4. During the call MS decrements the prepaid account which is stored locally in the SIM card.

*5) TOP Up System:*

Top up systems are used to update the prepaid balance. Traditional ways of top up are vouchers and rechargeable phone card. The problem with these methods is that user has to go to some store in order to buy more talking time. One more resent way to update the balance is by using cash machine[5], but this really does not solve the problem of physically going somewhere to update the balance. A more customer friendly and quite resent solution is to allow the customer to allocate one of her debit/credit cards for updating the balance[5]. This ways the customer just calls to operators call center whenever she wants to update the balance.

*c) Prepaid in GPRS networks*

Mobile Internet means Internet services provided via GPRS networks. Technical problem are more or less similar to problems that where experienced when prepaid was implemented in GSM networks. Solving these problems require acquisition of equipment that can handle real time billing and top upping of the prepaid account.

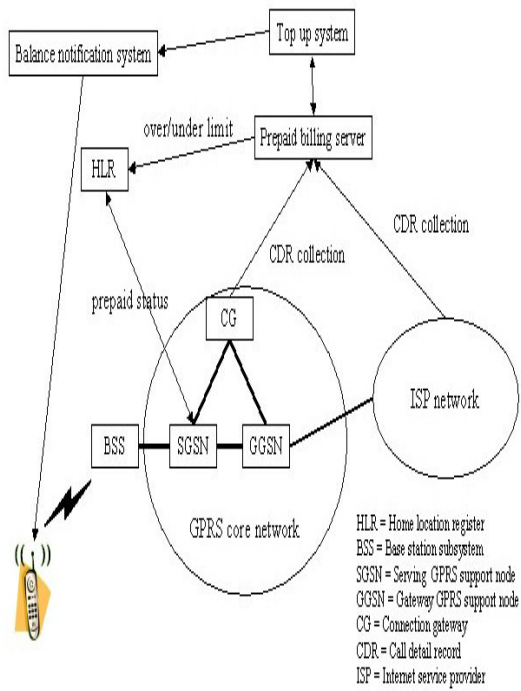

The connection procedure is much similar as the procedure in previous sections. The main difference in GPRS prepaid is that call detail records are gathered from both GPRS core network and ISP network. this adds to the complexity of GPRS prepaid system.

## III SECURITY ISSUES

➢ It is difficult to get information about customers.
➢ Customer loyalty is hard to maintain. Churn is higher with prepaid customers.
➢ Criminals prefer prepaid because of the anonymity.
➢ Real time charging systems are needed Operator needs a separate top up system.

## IV COUNTERMEASURES

According to the survey conducted in year 2012 under the umbrella of Nation Conference on Trade and Development revealed that in the Kingdom there are 188 registered mobile phones for every 100 users. The next is Macau in China, which for every 100 residents have 206 mobile phones.

To Avoid misuse of SIM cards ID number should be entered, together with a prepaid card number, would be made mandatory to charge a cell phones credit or to transfer money from the users credit that of another subscriber.

This procedure aims to ends the practice of anonymous persons misusing SIM cards

In this new system, a consumer should enter his ID immediately after entering the prepaid card number to recharge the phones credit.

The information supplied by subscribers while buying a SIM card must be correct and updated. It would not be possible to charge the credit if the users do not enter ID that they have finished while subscribing to the service.

## V CONCLUSION

This study explains that different technique of the charging a mobile phone prepaid credits. It shows that the existing schemes do not provide adequate security and that there is a need to develop new mechanism. It point out that anonymous SIM cards users were posing a server threat to the security of the nation and the society as well.
It Introduce a new technique to be reiterated to stop illegal SIM users. To implements this telecommunication companies should make the necessary changes in their system to support the regulation.

**Arshiya Begum** received the MCA degree and M.Tech degree in Computer Science from Jawaharlal Nehru Technological University, Andhra Pradesh, Hyderabad, India. She is Assistant professor in KSA.

**Mohammed Tanveer Ali** has done B.Tech in Electronic and Communication Engineering from Jawaharlal Nehru Technological University. He is having 7 years of Industrial Experience in Networking platform and Presently he is working in Computer Science Corporation as Network engineer.